\documentclass[aps,prl,twocolumn,preprintnumbers,%superscriptaddress,
              showpacs,nofootinbib]{revtex4}
\newcommand{\PRE}[1]{}       % Use if journal style
\usepackage{bm} \usepackage{epsfig}

\newcommand{\gweak}{g_{\text{weak}}}
\newcommand{\mweak}{m_{\text{weak}}}

\newcommand{\mev}{\text{MeV}}
\newcommand{\gev}{\text{GeV}}
\newcommand{\tev}{\text{TeV}}
\newcommand{\pb}{\text{pb}}

\newcommand{\cm}{\text{cm}}

\newcommand{\s}{\text{s}}

\newcommand{\etal}{{\em et al.}}
\newcommand{\eg}{{\em e.g.}}

\newcommand{\eqref}[1]{Eq.~(\ref{#1})}

\newcommand{\figref}[1]{Fig.~\ref{fig:#1}}

\newcommand{\mmess}{M_{\text{m}}}

\newcommand{\ssection}[1]{{\em #1.\ }}

\begin{document}

\preprint{UCI-TR-2008-10}

\title{
\PRE{\vspace*{1.5in}}
The WIMPless Miracle: Dark Matter Particles \\
without Weak-scale Masses or Weak Interactions
\PRE{\vspace*{0.3in}}
}

\author{Jonathan L.~Feng}
\affiliation{Department of Physics and Astronomy, University of
California, Irvine, CA 92697, USA
\PRE{\vspace*{.5in}}
}

\author{Jason Kumar%
\PRE{\vspace*{.2in}}
}
\affiliation{Department of Physics and Astronomy, University of
California, Irvine, CA 92697, USA
\PRE{\vspace*{.5in}}
}

%\date{March 2008}

\begin{abstract}
\PRE{\vspace*{.3in}} We propose that dark matter is composed of
particles that naturally have the correct thermal relic density, but
have neither weak-scale masses nor weak interactions.  These WIMPless
models emerge naturally from gauge-mediated supersymmetry breaking,
where they elegantly solve the dark matter problem. The framework
accommodates single or multiple component dark matter, dark matter
masses from 10 MeV to 10 TeV, and interaction strengths from
gravitational to strong.  These candidates enhance many direct and
indirect signals relative to WIMPs and have qualitatively new
implications for dark matter searches and cosmological implications
for colliders.
\end{abstract}

%95.35.+d Dark matter
%04.65.+e Supergravity
%12.60.Jv Supersymmetric models

\maketitle

\ssection{Introduction} Cosmological observations require dark matter
that cannot be composed of any of the known particles.  At the same
time, attempts to understand the weak force also invariably require
new states.  These typically include weakly-interacting massive
particles (WIMPs) with masses around the weak scale $\mweak \sim
100~\gev - 1~\tev$ and weak interactions with coupling $\gweak \simeq
0.65$.  An appealing possibility is that one of the particles
motivated by particle physics simultaneously satisfies the needs of
cosmology.  This idea is motivated by a striking quantitative fact,
the ``WIMP miracle'': WIMPs are naturally produced as thermal relics
of the Big Bang with the densities required for dark matter.  This
WIMP miracle drives most dark matter searches.

We show here, however, that the WIMP miracle does not necessarily
imply the existence of WIMPs.  More precisely, we present
well-motivated particle physics models in which particles naturally
have the desired thermal relic density, but have neither weak-scale
masses nor weak force interactions.  In these models, dark matter may
interact very weakly or it may couple more strongly to known
particles. The latter possibility implies that prospects for some dark
matter experiments may be greatly enhanced relative to WIMPs, with
search implications that differ radically from those of WIMPs.

Quite generally, a particle's thermal relic density
is~\cite{Zeldovich:1965}
\begin{equation}
\label{omega}
\Omega_X \propto {1\over \langle \sigma v \rangle}
\sim \frac{m_X^2}{g_X^4} \ ,
\end{equation}
where $\langle \sigma v \rangle$ is its thermally-averaged
annihilation cross section, $m_X$ and $g_X$ are the characteristic
mass scale and coupling entering this cross section, and the last step
follows from dimensional analysis.  In the models discussed here,
$m_X$ will be the dark matter particle's mass.  The WIMP miracle is
the statement that, for $(m_X, g_X) \sim (\mweak, \gweak)$, the relic
density is typically within an order of magnitude of the observed
value, $\Omega_X \approx 0.24$.  Equation (\ref{omega}) makes clear,
however, that the thermal relic density fixes only one combination of
the dark matter's mass and coupling, and other values of $(m_X,
g_X)$ can also give the correct $\Omega_X$.  Here, however, we further
show that simple models with low-energy supersymmetry (SUSY) predict
exactly the combinations of $(m_X, g_X)$ that give the correct
$\Omega_X$.  In these models, $m_X$ is a free parameter. For $m_X \ne
\mweak$, these models are WIMPless, but for all $m_X$ they contain
dark matter with the desired thermal relic density.

\begin{figure}
\resizebox{3.25in}{!}{
\includegraphics{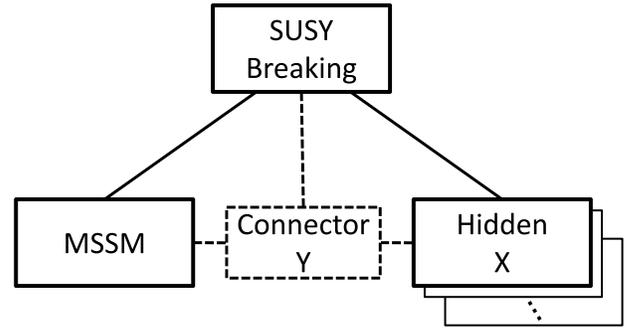}
}
\caption{Sectors of the model.  SUSY breaking is mediated by gauge
interactions to the MSSM and the hidden sector, which contains the
dark matter particle $X$.  An optional connector sector contains
fields $Y$, charged under both MSSM and hidden sector gauge groups,
which induce signals in direct and indirect searches and at colliders.
There may also be other hidden sectors, leading to multi-component
dark matter.
\label{fig:sectors}
\vspace*{-.2in}
}
\end{figure}

\ssection{Models} We will consider SUSY models with gauge-mediated
SUSY breaking (GMSB)~\cite{Dine:1981za,Dine:1994vc}.  These models
have several sectors, as shown in \figref{sectors}.  The MSSM sector
includes the fields of the minimal supersymmetric standard model.  The
SUSY-breaking sector includes the fields that break SUSY dynamically
and mediate this breaking to the MSSM through gauge interactions.
There are also one or more additional sectors which have SUSY breaking
gauge-mediated to them; these sectors contain the dark matter
particles.  These sectors may not be very well-hidden, depending on
the presence of connector sectors (discussed below), but we will
follow precedent and refer to them as ``hidden'' sectors. For other
recent studies of hidden dark matter, see Refs.~\cite{hiddenDM}.

This is a well-motivated scenario for new physics.  GMSB models
feature many of the virtues of SUSY, while elegantly solving the
flavor problems that generically plague proposals for new weak-scale
physics.  Additionally, in SUSY models that arise from string theory,
hidden sectors are ubiquitous.  As a concrete example, we extend the
canonical GMSB models of Ref.~\cite{Dine:1994vc} to include one hidden
sector.  SUSY breaking gives vacuum expectation values to a chiral
field $S$, with $\langle S \rangle = M + \theta^2 F$.  We couple $S$
to MSSM messenger fields $\Phi$, $\bar{\Phi}$ and hidden sector
messenger fields $\Phi_X$, $\bar{\Phi}_X$ through the superpotential
$W = \lambda \bar{\Phi} S \Phi + \lambda_X \bar{\Phi}_X S \Phi_X$.
These couplings generate messenger $F$-terms $F_{\text{m}} = \lambda
F$ and $F_{\text{m} X} = \lambda_X F$ and induce SUSY-breaking masses
in the MSSM and hidden sectors at the messenger mass scales $\mmess =
\lambda M$ and $M_{\text{m} X} = \lambda_X M$, respectively.

\ssection{Relic Density} Neglecting subleading effects and ${\cal
O}(1)$ factors, the MSSM superpartner masses are
\begin{equation}
\label{mmass}
m \sim \frac{g^2}{16 \pi^2} \frac{F_{\text{m}}}{M_{\text{m}}}
= \frac{g^2}{16 \pi^2} \frac{F}{M} \ ,
\end{equation}
where $g$ is the largest relevant gauge coupling.  Since $m$ also
determines the electroweak symmetry breaking scale, $m \sim \mweak$.
The hidden sector superpartner masses are
\begin{equation}
\label{mxmass}
m_X \sim \frac{g_X^2}{16 \pi^2} \frac{F_{\text{m} X}}{M_{\text{m} X}}
= \frac{g_X^2}{16 \pi^2} \frac{F}{M} \ .
\end{equation}
As a result,
\begin{equation}
\frac{m_X}{g_X^2} \sim \frac{m}{g^2} \sim
\frac{F}{16 \pi^2 M} \ ;
\label{mxgx}
\end{equation}
that is, $m_X/g_X^2$ is determined solely by the SUSY-breaking sector.
As this is exactly the combination of parameters that determines the
thermal relic density of \eqref{omega}, the hidden sector
automatically includes a dark matter candidate that has the desired
thermal relic density, irrespective of its mass.  (In this example,
the superpartner masses are independent of $\lambda$ and $\lambda_X$;
this will not hold generally.  However, given typical couplings
$\lambda \sim \lambda_X \sim {\cal O}(1)$, one expects the messenger
$F$-terms and masses to be approximately the same as those appearing
in $\langle S \rangle$, and \eqref{mxgx} remains valid.)

This analysis assumes that these thermal relics are stable.  Of
course, this is not the case in the MSSM sector, where thermal relics
decay to gravitinos.  This is a major drawback for GMSB, especially
because its classic dark matter candidate, the thermal
gravitino~\cite{Pagels:1981ke}, is now too hot to be compatible with
standard cosmology~\cite{Seljak:2006qw}.  Solutions to the dark matter
problem in GMSB include messenger sneutrinos~\cite{Han:1997wn}, late
entropy production~\cite{Baltz:2001rq}, decaying
singlets~\cite{Ibe:2006rc}, and gravitino production in late
decays~\cite{Feng:2008zz}, but all of these bring complications, and
only the last one makes use of the WIMP miracle.

But the problem exists in the MSSM only because of an accident: the
stable particles of the MSSM ($p$, $e$, $\nu$, $\gamma$, $\tilde{G}$)
have masses which are not at the scale $\mweak$.  For the proton and
electron, this accident results from extremely suppressed Yukawa
couplings which are unexplained.  There is no reason for the hidden
sector to suffer from this malady.  Generally, since $m_X$ is the only
mass scale in the hidden sector, we expect all hidden particles to
have mass $\sim m_X$ or be essentially massless, if enforced by a
symmetry.  We assume that the thermal relic has mass around $m_X$, and
that discrete or continuous symmetries stabilize this particle.  The
particles that are essentially massless at freeze out provide the
thermal bath required for the validity of \eqref{omega}.  An example
of a viable hidden sector is one with MSSM-like particle content (with
possible additional discrete symmetries), but with different gauge
couplings and with all Yukawa couplings ${\cal O}(1)$.  The light
particles are then the neutrinos, gluon, photon (and gravitino), while
the remaining particles are all at the scale $m_X$.  The lightest such
particle charged under a (possibly discrete) unbroken symmetry will
then be stable by hidden sector charge conservation.

One might worry that the extra light particles will have undesirable
cosmological consequences.  In particular, the number of light
particles are constrained by Big Bang nucleosynthesis
(BBN)~\cite{Cyburt:2004yc} and (less stringently) the cosmic microwave
background~\cite{Komatsu:2008hk} even if they have no SM interactions.
These constraints have been analyzed in detail in
Ref.~\cite{Feng:2008mu}.  They are found to require $g_*^h
(T^h_{\text{BBN}} / T_{\text{BBN}} )^4 \le 2.52 \ \text{(95\% CL)}$,
where $g_*^h$ is the number of relativistic degrees of freedom in the
hidden sector at BBN, and $T^h_{\text{BBN}}$ and $T_{\text{BBN}}$ are
the temperatures of the hidden and observable sectors at BBN,
respectively.  This bound may therefore be satisfied if, for example,
$g_*^h < 2.5$ or if the hidden sector is as big as the MSSM with
$g_*^h = 10.75$ but is slightly colder, with $T^h_{\text{BBN}} /
T_{\text{BBN}} < 0.7$.  Such discrepancies in temperature are possible
if the observable and hidden sectors reheat to different
temperatures~\cite{Hodges:1993yb,Berezhiani:1995am} and need not alter
the relic density calculation significantly~\cite{Feng:2008mu}.

To summarize so far: GMSB models with hidden sectors provide dark
matter candidates that are not WIMPs but nevertheless naturally have
the correct thermal relic density.  These candidates have masses and
gauge couplings satisfying $m_X/g_X^2 \sim \mweak / \gweak^2$, and
\begin{eqnarray}
10^{-3} \alt \! &g_X& \! \alt 3  \nonumber \\
10~\mev \alt \! &m_X& \! \alt 10~\tev \ ,
\end{eqnarray}
where the upper limits from perturbativity nearly saturate the
unitarity bound~\cite{Griest:1989wd}, and the lower limits are rough
estimates from requiring the thermal relic to be non-relativistic at
freeze out so that \eqref{omega} is valid.

\ssection{Detection} If the hidden sector is not directly coupled to
the SM, then the corresponding dark matter candidate interacts with
the known particles extremely weakly.  A more exciting possibility
is that dark matter interactions are
enhanced by connector sectors containing particles $Y$ that are
charged under both MSSM and the hidden sector, as shown in
\figref{sectors}.

$Y$ superpartner masses receive contributions from both MSSM and
hidden sector gauge groups, and so we expect $m_Y \sim \max ( \mweak,
m_X)$.  Connectors interact through $\lambda XYf$, where $\lambda$ is
a Yukawa coupling and $f$ is a SM particle. $X$ remains stable, as
long as $m_X < m_Y + m_f$, but these interactions mediate new
annihilation processes $X \bar{X} \to f \bar{f}, Y \bar{Y}$ and
scattering processes $X f \to X f$.  The new annihilation channels do
not affect the thermal relic density estimates given above, provided
$\lambda \alt \gweak$.

Connector particles create many new possibilities for dark matter
detection.  For example, in WIMPless models, the dark matter may have
$m_X \ll \mweak$.  This motivates direct searches probing masses far
below those typically expected for WIMPs.  Because the number density
must compensate for the low mass, indirect detection signals are
enhanced by $\mweak^2/m_X^2$ over WIMP signals.

To quantify this, we consider a simple connector sector with chiral
fermions $Y_{f_L}$ and $Y_{f_R}$ and interactions
\begin{equation}
{\cal L} = \lambda_f X \bar{Y}_{f_L} f_L
+ \lambda_f X \bar{Y}_{f_R} f_R
+ m_{Y_f} \bar{Y}_{f_L} Y_{f_R} \ ,
\label{connector}
\end{equation}
where the fermions $f_L$ and $f_R$ are SM SU(2) doublets and singlets,
respectively.  The $Y_f$ particles get mass from SM electroweak
symmetry breaking.  For simplicity, we couple $Y$ to one SM particle
$f$ at a time, but, one $Y$ can have multiple couplings or there can
be many $Y$ fields.

We begin with direct detection, and assume the interactions of
\eqref{connector} with $f = u$.  These mediate spin-independent
$X$-nucleus scattering through $X u_{L,R} \to Y_{L,R} \to X u_{L,R}$
with cross section
\begin{equation}
\sigma_{\text{SI}} = \frac{\lambda_u^4}{2\pi}
\frac{m_N^2}{(m_N + m_X)^2}
\frac{\left[ Z B^p_u + (A-Z) B^n_u \right]^2}
{(m_X - m_Y)^2} \ ,
\end{equation}
where $A$ ($Z$) is the atomic mass (number) of nucleus $N$, $B^p_u =
\langle p | \bar{u} u | p \rangle \simeq 5.1$, and $B^n_u = \langle n
| \bar{u} u | n \rangle \simeq 4.3$~\cite{Cheng:1988im}.

\begin{figure}
\resizebox{3.25in}{!}{
\includegraphics{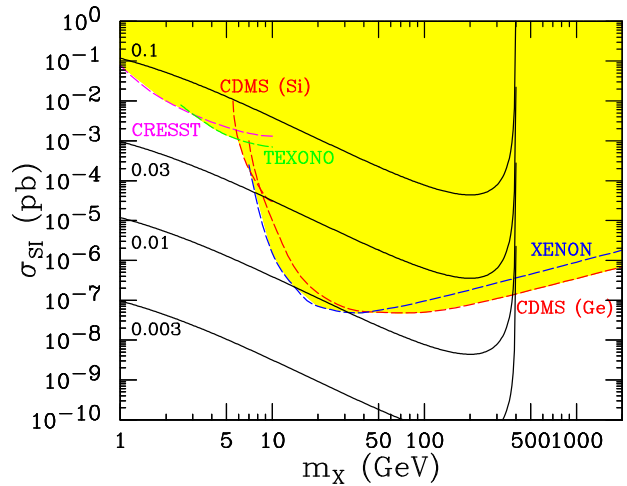}
}
\caption{Direct detection cross sections for spin-independent
$X$-proton scattering as a function of dark matter mass $m_X$.  The
solid curves are the predictions for WIMPless dark matter with
connector mass $m_{Y_u}=400~\gev$ and the Yukawa couplings $\lambda_u$
indicated.  The shaded region is excluded by
CRESST~\cite{Angloher:2002in}, CDMS (Si)~\cite{Akerib:2005kh},
TEXONO~\cite{Lin:2007ka}, XENON~\cite{Angle:2007uj}, and CDMS
(Ge)~\cite{Ahmed:2008eu}.
\label{fig:direct}
\vspace*{-.2in}
}
\end{figure}

In \figref{direct}, we present $X$-proton scattering cross sections as
functions of $m_X$ for various $\lambda_u$ and $m_{Y_u} = 400~\gev$.
$Y_u$ receives mass from SM electroweak symmetry breaking, and this
mass is well within bounds from perturbativity and experimental
constraints~\cite{Kribs:2007nz}.  Note that the cross sections are
much larger than for neutralinos and many standard WIMPs, such as
$B^1$ Kaluza-Klein dark matter~\cite{Cheng:2002ej}.  Also, the
framework accommodates dark matter at the GeV or TeV scale, which may
resolve current anomalies, such as the apparent conflict between DAMA
and other experiments~\cite{Gondolo:2005hh}.

We now turn to indirect detection and consider the interactions of
\eqref{connector} with $f = \tau$.  These interactions can
produce excess photon fluxes from the galactic center.  The
integrated flux is~\cite{Feng:2000zu}
\begin{equation}
\Phi_{\gamma} =
\frac{5.6 \times 10^{-10}}{\cm^2~\s}
N_{\gamma} {\sigma_{\text{SM}} v \over \pb}
\left[{100~\gev \over m_X}\right]^2
\bar{J} \Delta \Omega \ ,
\end{equation}
where the cross section for $X \bar{X}
\to \tau^+ \tau^-$ is
\begin{equation}
\sigma_{\text{SM}} v = {\lambda_{\tau}^4 \over 4\pi}
{m_Y ^2 \over (m_X ^2 +m_Y ^2)^2} \ ,
\end{equation}
${\bar J}$ is a constant parameterizing the cuspiness of our galaxy's
dark matter halo, $\Delta \Omega$ is the experiment's solid angle, and
$N_{\gamma} = \int_{E_{\text{thr}}}^{m_X} dE {dN_{\gamma} \over dE}$
is the average number of photons above threshold produced in each
$\tau$ decay.

\begin{figure}
\resizebox{3.25in}{!}{
\includegraphics{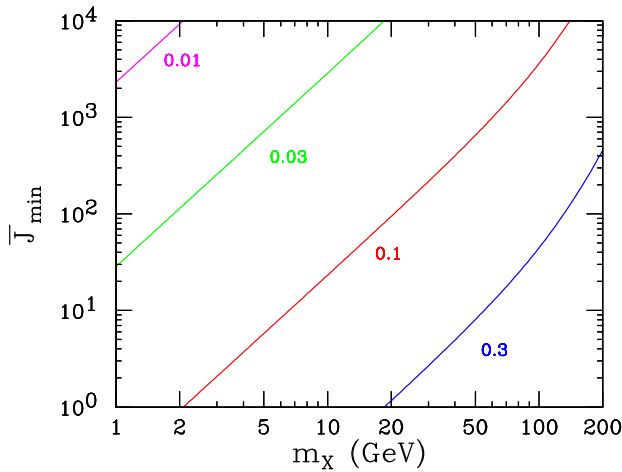}
}
\caption{Indirect detection prospects for WIMPless dark matter as a
function of dark matter mass $m_X$.  For values of $\bar{J}$ above the
contours, the annihilation process $X \bar{X} \to \tau \bar{\tau}$
yields an observable photon signal at GLAST.  We assume connector mass
$m_{Y_{\tau}} = 200~\gev$ and the Yukawa couplings $\lambda_{\tau}$
indicated.
\label{fig:indirect}
\vspace*{-.2in}
}
\end{figure}

In \figref{indirect}, we evaluate the discovery prospects for
GLAST~\cite{GLAST}.  We take $\Delta \Omega = 0.001$, $N_{\gamma} = 1$
and $E_{\text{thr}} = 1~\gev$, and require $\Phi_{\gamma} >
10^{-10}~\cm^{-2}~\s^{-1}$ for discovery.  The minimum values of
$\bar{J}$ for discovery for various $\lambda_{\tau}$ as a function of
$m_X$ are given in \figref{indirect}. As the flux is proportional to
number density squared, we find excellent discovery prospects for
light dark matter.  For $\lambda_{\tau} =0.3$ and $m_X \alt 20~\gev$,
GLAST will see WIMPless signals for ${\bar J} \sim 1$, corresponding
to smooth halo profiles that are inaccessible in standard WIMP models.

\ssection{Conclusions} In GMSB models with hidden sectors, we have
found that, remarkably, any stable hidden sector particle will
naturally have a thermal relic density that approximately matches that
observed for dark matter.  Indeed, it is merely an accident that the
MSSM itself has no stable particle with the right relic density in
GMSB, and it is an accident that need not occur in hidden sectors.
These candidates possess all the key virtues of conventional WIMPs,
but they generalize the WIMP paradigm to a broad range of masses and
gauge couplings.  This generalization opens up new possibilities for
large dark matter signals.  We have illustrated this with two
examples, but many other signals are possible.

As shown in \figref{sectors}, this scenario also naturally
accommodates multi-component dark matter if there are multiple hidden
sectors.  This is highly motivated --- in IBMs, one generally expects
multiple hidden sectors in addition to the MSSM.  In this framework,
it is completely natural for dark matter particles with varying masses
and couplings to each be a significant component of dark matter.

Finally, WIMPless dark matter introduces new possibilities for the
interplay between colliders and dark matter searches.  For example,
LHC evidence for GMSB would exclude neutralino dark matter, but favor
WIMPless (and other) scenarios. Further evidence from direct and
indirect searches, coupled with Tevatron or LHC discoveries of ``4th
generation'' quarks or leptons, could disfavor or establish the
existence of WIMPless dark matter and the accompanying connector
sectors.

\ssection{Acknowledgments} We gratefully acknowledge
S.~Palomares-Ruiz, A.~Pierce, A.~Rajaraman, R.~Schnee, Y.~Shirman, and
X.~Tata for useful discussions. This work was supported in part by NSF
grants PHY--0239817, PHY--0314712, and PHY--0653656, NASA grant
NNG05GG44G, and the Alfred P.~Sloan Foundation.

%%%%%%%%%%%%%%%%%%%%%%%%%%%%%%%%%%%%%%%%%%%%%%%%%%%%%%

%%%%%%%%%%%%%%%%%%%%%%%%%%%%%%%%%%%%%%%%%%%%%%%%%%%%%%

\end{document}